\RequirePackage{fix-cm}
\documentclass[smallextended]{svjour3}       
\smartqed  
\usepackage{graphicx}
\usepackage{comment}
\usepackage{mathtools}

\DeclarePairedDelimiter\ket{\lvert}{\rangle}
\DeclarePairedDelimiterX\braket[2]{\langle}{\rangle}{#1 \delimsize\vert #2}

\begin{document}

\title{Continuous-Variable Deep Quantum Neural Networks for Flexible Learning of Structured Classical Information
}


\author{Jasvith Raj Basani         \and
        Aranya Bhattacherjee 
}


\institute{Jasvith Raj Basani \at
              Department of Electrical and Electronics Engineering, Birla Institute of Technology and Science,
				Pilani, Hyderabad Campus, Telangana, 500078, India \\
              Tel.: +(91) 9986546980\\
              \email{f20170374@hyderabad.bits-pilani.ac.in}           
           \and
           Aranya Bhattacherjee \at
              Department of Physics, Birla Institute of Technology and Science,
				Pilani, Hyderabad Campus, Telangana, 500078, India
}

\date{Received: date / Accepted: date}

\maketitle

\begin{abstract}

Quantum computation using optical modes has been well-established in its ability to construct deep neural networks. These networks have been shown to be flexible both architecturally as well as in terms of the type of data being processed. We leverage this property of the Continuous-Variable (CV) model to construct stacked single mode networks that are shown to learn structured classical information, while placing no restrictions on the size of the network, and at the same time maintaining it's complexity. The hallmark of the CV model is its ability to forge non-linear functions using a set of gates that allows it to remain completely unitary. The proposed model exemplifies that the appropriate photonic hardware can be integrated with present day optical communication systems to meet our information processing requirements. In this paper,  using the Strawberry Fields software library on the MNIST dataset of hand-written  digits,  we  demonstrate  the  adaptability  of  the  network  to  learn  classical  information  to fidelities of greater than 99.98\%.

\end{abstract}

\section{Introduction}
\label{intro}

Neural  networks  enjoy  widespread  success  both  in academia  and  the  industry,  with  a  plethora  of  applications  ranging  from  solving  simple  classification  problems to information security and processing. The review article by Pal and Pal \cite{1} on image segmentation predicted that neural networks would have a multitude of applications even in the field of image processing. One of the most representative models - the Deep Convolutional Neural Network, has been widely used in diverse areas from time-dependent signal processing to computer vision. However, despite these plethora of applications, the dramatic increase in the computational cost has been deemed as a bottleneck in the functionality of these networks. Quantum computing, however, seems to provide a viable alternative to satisfy our growing demand for computational power and efficiency.

The advent of quantum technologies has resulted in what is being referred to as quantum-enhanced machine learning where we expect a speed-up either by employing genuine quantum effects, or by classical machine learning to improve quantum processes.  A hybrid classical-quantum system achieves  this  speed-up  by  outsourcing  computationally difficult subroutines to the integrated quantum device - specifically the quadratures of light in a quantum optical system. The KLM protocol \cite{2}, proposes that quantum computing can be achieved solely using linear-optic tools, single photon sources, photodetectors and ancilla resources. In the following paper, we demonstrate the use of the Continuous-Variable (CV) model to flexibly learn structured data without restrictions on the format of the data, while proposing a method to realize the same experimentally via parameter-dependent gates. 

Information systems have always had the need for components that would effectively be able to handle a multitude of signals and could easily be integrated in existing networks. The following work introduces such an element that we show to be adaptable to our signal processing requirements by means of numerical experiments with some of the most fundamental applications of deep learning networks on the platform of image processing - namely image classification, image reconstruction and image enhancement. The versatility of the kind of information being learn is illustrated in these cases with the architecture modified to suit to specifics of the data. Note that while an image could be considered as a two-dimensional signal, there is no feature of the network that confines it to learn such data - higher dimensional tensors and time-dependent signals can be learnt equally well with the appropriate datasets, by simply stacking the proposed layers one after the other. Furthermore, we propose an experimental model that is commercially feasible and can be assimilated into modern optical communication systems.

This paper is organized as follows - in section 2, be provide a brief theoretical background to the CV model as well introduce the construction for the experimental implementation of our network. We then demonstrate the results of our numerical simulations for fundamental applications in image processing. Finally, after a discussion of the merits of the model, we conclude with the applications and future scope in this field.

\section{Continuous-Variable Single Mode Quantum Layer}

\subsection{Operating Principles of the Quantum Layer}

The widely accepted model of the qubit based quantum computer has proven to be ill-suited to tackle continuous-valued problems \cite{3}.  Fortunately, the CV  model \cite{4} proposes  an  alternative  to  the  qubit  -  the qumode, wherein information is encoded in the quantum states  of  bosonic  modes  -  the  ubiquity  and  features  of which will be further discussed. As opposed to the conventional discrete set of coefficients which are used in the qubit expansions, the CV model uses a continuum of states as seen by the expansion of a qumode below: 

\begin{eqnarray*}
\ket{\psi} = \int    \psi(x) \ket{x} dx
\end{eqnarray*}
where $\ket{x}$ are the eigenstates of the $\hat{x}$ quadrature, which will further be elaborated upon below. In the following sections, we reference the phase space formalism of the continuous variable quantum mechanics where we treat the conjugate variables $x$ and $p$ on equal footing which allows us to find symmetries with classical Hamiltonian dynamics. The universality of the CV model has been well established and is based upon its ability to approximate a broad set of Hamiltonians that are polynomials, fixed in degree, which are functions of the $\hat{x}$ and $\hat{p}$ quadratures. In the discrete qubit system, we are enabled by a set of gates that allow any normalized system to transform through unitary operations, while in the CV model, we apply Gaussian and Non-Gaussian transformations to evaluate the the evolution of a state that takes the form $\ket{\psi} = e^{-i \hat{H}t}\ket{0}$ where $\hat{H}$ is the generator or the Hamiltonian of the bosonic system and $\ket{0}$ is the vacuum state which we initialize to assume time dependant evolution.

Now that we have established that the CV model can be used for solving continuous-variable problems such as those generally solved by neural networks, we can proceed with the design of a single quantum layer. The output of the network can be described as the following :
\begin{eqnarray*}
\ket{\textbf{x}} \rightarrow \ket{\sigma(W\textbf{x} + \textbf{b})}
\end{eqnarray*}
From figure 1, we see the implementation of a layer L of the CV based quantum neural network as proposed in \cite{5}. As apparently presented, the function of the layer can be broken down as follows : 
\begin{eqnarray}
L = \Phi\cdot D \cdot U_{2} \cdot S \cdot U_{1}
\end{eqnarray}
Here, $U_{i}$ is the $i^{th}$ N mode interferometer, that can further be decomposed rectangularly or triangularly into combinations of beamsplitters and rotation gates, hence is a function of the polarization angle as well as the azimuthal rotation angle. Furthermore, $S$ is the single mode squeeze gate, $D$ is the single mode displacement gate and $\Phi$ is the Non-Gaussian gate acting on a single mode, acting as non-linear activation function. The model has been posited in [5], where the working can been covered in depth. The linear section of the model has been achieved via the singular value decomposition \cite{6} of the matrix $W$ into rotation and squeezing operations, followed by a displacement operation to introduce the bias term, followed by the Non-Gaussian gate for the non-linear activation function. 

Deep learning models \cite{7} are made by stacking layers of the aforementioned circuit end-to-end while varying the number of qumodes in each layer for the based on the amount of information encoded into each qumode.

\subsection{Linear Optical Realization of the Quantum Layer}

The CV formalism has long been established as an experimentally realizable and flexible scheme to operate upon, with previous literature covering optical systems \cite{8,9}, ion traps \cite{10,11} and microwave systems \cite{12,13}. We illustrate a method to realize the architecture proposed experimentally using only linear-optic components. One must note here that the gates should be made tunable or be programmed from with values from numerical simulations. For optical hardware that is completely tunable, the gate parameters will need to be updated via control signals fed from the computer running the optimization algorithm. 

The first unit, which is the Displacement gate is used to initialize the photon with the $\hat{x}$ and $\hat{p}$ quadrature values so as to feed the quantum layers with the encoded information. The same has been achieved experimentally \cite{14} using a Beam-Splitter with the inputs as the photon to be initialized and a highly excited coherent state. The theory for the construction of a Displacement gate shows us that irrespective of the basis of calculation used, a single beam-splitter serves our purpose. The displacement gate used in the quantum layer can be constructed similarly. Furthermore, the coherent state is tuned using feedback from the laser, which can be solved for from the output of the Beamsplitter via Heisenberg's field equations. 

A careful analysis of the function of the interferometer reveals that while operating on a single mode, it functions as a rotation gate and hence the interferometer is extraneous. The construction of both interferometers and programmable phase shifters are very well-established. The next gate, for squeezing is a commonly used operation, with previous literature showing upwards of 15dB of squeezed light for applications in wave interferometry \cite{15}. While this is conventionally performed on an integrated photonic chip, it has also been performed via linear optic elements as demonstrated in \cite{16}. This combination of the Interfereometers and the Squeeze gate allows us to perform linear matrix operations on the mode via singular value decomposition. Finally, the nonlinear component, propounded in as the Kerr gate can be replaced in principle by either quadratic or cubic phase gates, both of which can be achieved by using either noncentrosymmetric or centrosymmetric materials - to be decided by the function of the network - in an Optical Parametric Oscillator.

The simplicity and the feasibility of this model is the key takeaway from this section, where any changes in the model can be performed by simply adding gates and changes in the architecture can be done by stacking such layers one after the other to structure the network as required. We use this feature for image processing functions below. 

\begin{figure*}
\centering
\includegraphics[scale = 0.7]{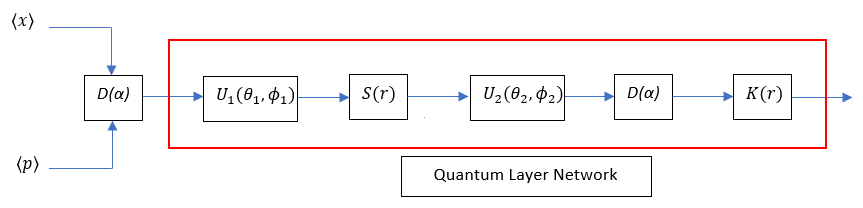}
\caption{\label{fig:wide} Single Mode Quantum Decoder Layer}
\end{figure*}

In the context of a conventional deep neural network, we shall refer to the number of layers as the depth of the network and the number of neurons per layer as the size of the layer.

\section{Classification}

In the following section, we demonstrate one of the most cardinal functions of neural networks in image processing - classifying the MNIST dataset \cite{17}. Here, we use an optical setup, with the image being encoded classically into multiple modes learning probabilistic values, each of which passes through a single layer quantum circuit as elucidated above. The images from the MNIST dataset are grayscale, with dimensions of 28 pixels by 28 pixels. Each image is reshaped into a 784 x 1 vector and reduced into a 20 x 1 vector by means of an arbitrary matrix multiplication that acts as an encoding operation. Each of these 20 values represents the $\hat{x}$ or $\hat{p}$ quadratures of the 10 modes. Our network was designed to learn values only in the $\hat{x}$ quadrature, i.e. the true outputs - which we shall refer to as $\ket{\phi}$ were vectors with all the $\hat{p}$ quadrature values set to 0, and the corresponding $\hat{x}$ quadrature excited to 1.

The simplicity of the architecture of the network reflects its feasibility for construction as well as experimental realization.  Since this system deals with 10 modes, naturally, the size of the layer is 10 units. We demonstrate that with a single layer i.e. depth set to 1, we can achieve classification accuracies of 94.9\%. For experimental implementation, the size of the layer can be changed easily - as determined by the output parameters of the dataset being used. Furthermore, the control signals, beamsplitters as well as the coherent laser sources can be tuned based on the trainable variable values as returned by the numerical experiments. Our simulations were based on the output state of the network learning the true state which can be written probabilistically as the mean squared difference between the fidelity, denoted by  $\braket{\psi}{\phi}$ and unity: 

\begin{equation}
C = \Sigma_{i} ^ {10}( |\braket{\psi_{i}}{\phi_{i}}|^{2} - 1)^{2}
\end{equation}

We provide a contrast between the softmax outputs of the network when trained classically and when trained on the CV model, where we expect to see the highest probabilities of prediction along the main diagonal. The classical network still does present with higher probabilities of finding the correct outputs - this can be attributed to several factors, primarily the fact that the CV-based network is trained on the $\hat{x}$ quadrature, implying that some information is still lost to the $\hat{p}$ quadrature. Placing better constraints on the network, or introducing another Displacement gate at the end of each layer so as to set the $\hat{p}$ quadratures to 0 appears to be a feasible method of gaining even higher accuracies.

\begin{figure}[h]
\centering
\includegraphics[width = 9cm]{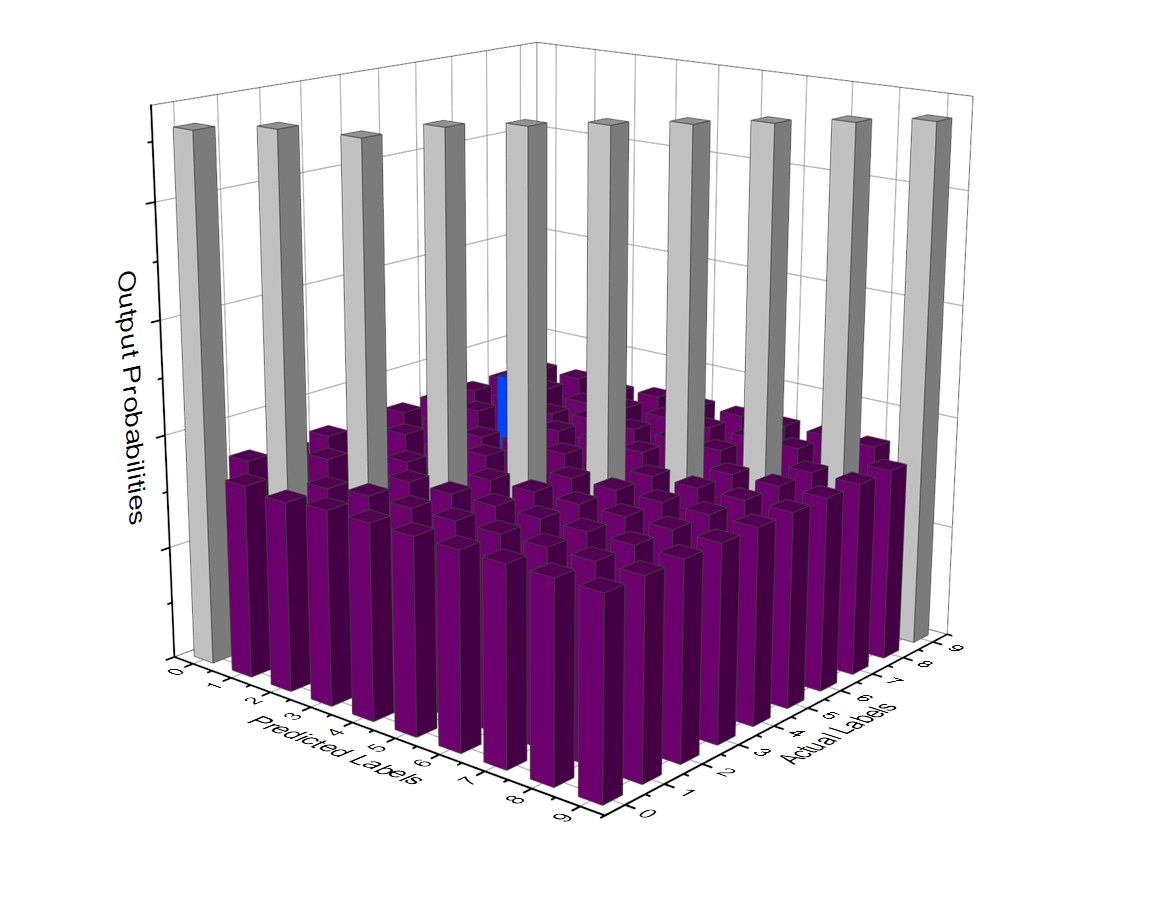}
\includegraphics[width = 9cm]{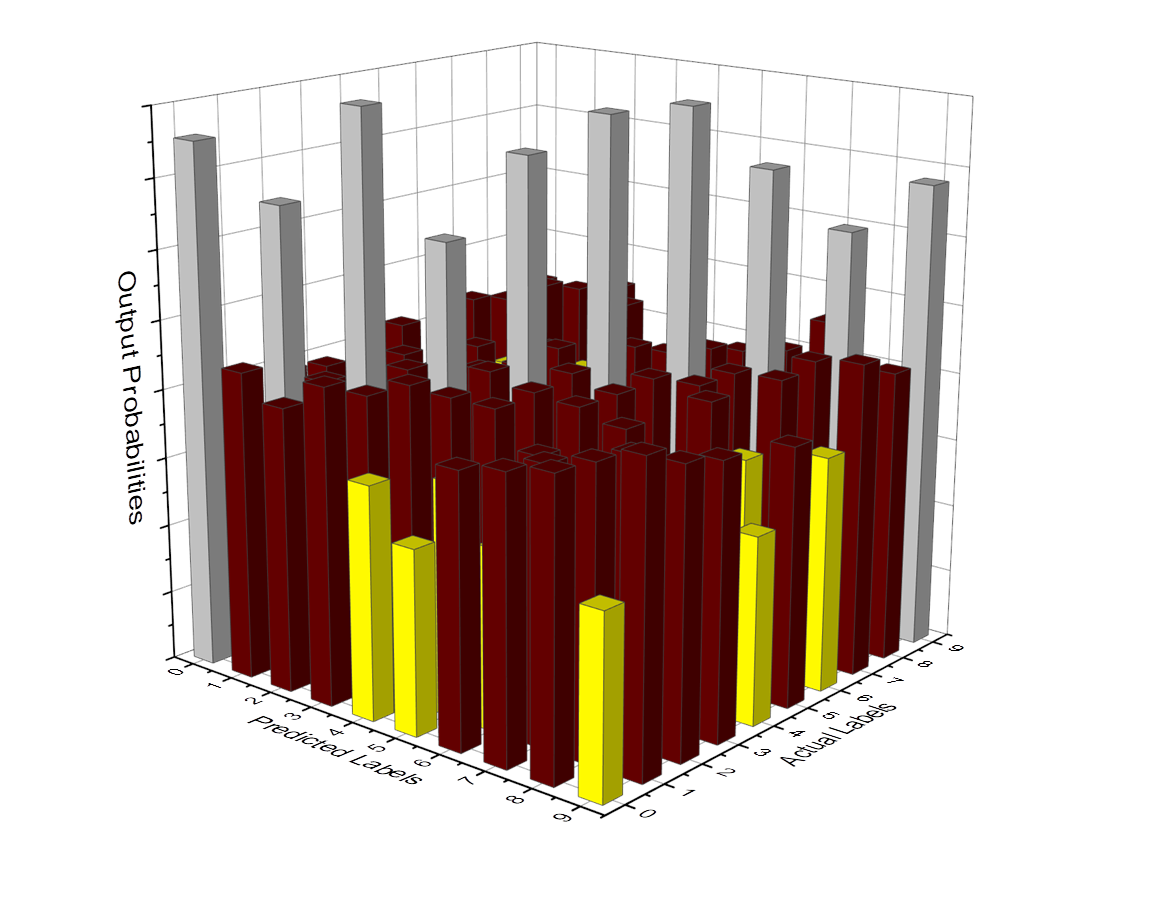}
\caption{\label{fig:epsart} Tensorflow softmax outputs of the network for classical and quantum classification of the MNIST dataset. The X and Y axes are the predicted values and the actual labels respectively}
\end{figure} 

Another salient function of neural networks when applied to classification problems is object recognition. This is, however, conventionally performed using convolutional neural networks, the single mode CV analogue of which remains to be explored in future work. Our model has performed the classification on the MNIST dataset with 10 single mode quantum layers. This function can be performed with 1 single mode layer by projecting the output of the network into higher dimensional image Hilbert space which we demonstrate in the following sections. Furthermore, since this technique places no restrictions on the type of data being used, with appropriate modifications in the architecture, similar designs could be used for classification of multi-dimensional signals as well, thereby reiterating its flexibility.

\section{Reconstructing an Encoded Image}

In this section, we demonstrate the use of a single mode in the 28-dimensional basis for the reconstruction of encoded images. The encoding is done classically, wherein via arbitrary matrix multiplications, we reduce the reshaped image into a 2 x 1 vector, each term representing the expectation values of the $\hat{x}$ and $\hat{p}$ quadratures to be passed as the first and second observables via the first displacement gate. Here, an important note to be made is that with a slight modification in labelling the observables, we can encode higher ($N$) dimensional information into a singe qumode - the observables being a set of matrices with the $N^{th}$ diagonal  element set to unity - giving us linear, hermitian and real-eigenvalued operators. 

Since we use only 1 mode in this model, our architecture uses only 1 neuron per layer, with a depth of 15 layers. Our first step is to define the mode upon which we shall be performing the learning operations. For this, we define the normalized projection as follows, with the projection performed onto the subspace of the first 28 Fock states. with $\Pi_{28}$ representing the corresponding projection operator. This operation is essential in ensuring that the model returns the state $\ket{\psi_{i}}$ upon successful projection into the subspace. 
\begin{eqnarray}
\ket{\psi_{i}} = \frac{\hat{\Pi}_{28}\ket{\Psi_{i}}}{||\hat{\Pi}_{28}\ket{\Psi_{i}}||}
\end{eqnarray}

Using techniques commonly used in Principle Component Analysis, we posit the fact that this projection can be generalized to an arbitrary dimensional space. Let us consider a vector $X$ with a 2-dimensional input ($\hat{x}$ and $\hat{p}$ from the encoding layer) which needs to be projected into the $R^{28}$ space as the vector $\Pi_{U}(X)$, where U represents the space spanned by the orthogonal basis vectors $B = \{ \textbf{b}_{1}, \textbf{b}_{2}, \textbf{b}_{3} . . . \textbf{b}_{28} \}$. Hence the vector $\Pi_{U}(X)$ can be written as a linear combination of the basis vectors as $\Pi_{U}(X) = \lambda B$, where $\lambda = \{\lambda_{1}, \lambda_{2}, \lambda_{3} . . . \lambda_{28}\}$.\\
Here, we take advantage of the orthogonality of the basis vectors and hence, can write $\braket{\Pi_{U}(X) - X}{B} = 0$. Simple algebra, while generalizing to the entire basis, gives us $\lambda^{T}B^{T}B - X^{T}B = 0$, when we replace $\Pi_{U}(X)$ with $\lambda B$. Hence, solving for $\lambda$ and $\Pi_{U}(X)$, we get: 
\begin{eqnarray*}
\lambda = (B^{T}B)^{-1}B^{T}X
\end{eqnarray*}
\begin{eqnarray*}
\Pi_{U}(X) = B\lambda = B(B^{T}B)^{-1}B^{T}X
\end{eqnarray*}

This projected state is then modified as the network learns from the cost function given below, with the "correctness" between the target state and the learnt state measured by the fidelity defined by $\braket{i}{\psi_{i}}$
\begin{eqnarray}
C = \Sigma_{i = 0}^{27} (|\braket{i}{\psi_{i}}|^2 - 1)^2 + \gamma P({\ket{\Psi_{i}}})
\end{eqnarray}

Implementing the above using the Strawberry Fields \cite{18,19} package available commercially, applied with the support of Tensorflow \cite{20}, gives us the reconstructed image as shown in the figures on the right in figure [3]. We have used both the Adam \cite{21} and RMSProp \cite{22} optimizers with the learning rate $\alpha$ set to 0.001.

\begin{figure}[h]
\centering
\includegraphics[width = 3cm]{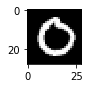}
\includegraphics[width = 3cm]{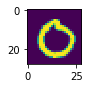}\\
\includegraphics[width = 3cm]{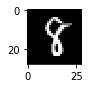}
\includegraphics[width = 3cm]{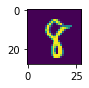}
\caption{\label{fig:epsart} Inputs (left) and outputs (right) of the hybrid net-work - the images of the numbers ’0’ and ’8’ from the MNIST dataset of handwritten digits}
\end{figure}

\section{Image Denoising and Enhancement}

In the following section, we explore the aspect of denoising the input image with Additive White Gaussian Noise (AWGN). The Central Limit Theorem \cite{23} establishes that the sum of independent random variables - in this case kinds of noise - tends towards a normal distribution, even if the original variables themselves are not normally distributed. AWGN is a model of noise, commonly used in information theory and processing so as to model the random effects of interactions that occur in nature. We leverage the property that this noise is white - meaning that the power spectrum of the noise is a constant and that it is Gaussian.

The proposed method for denoising uses a method similar to that described above, the difference being, in this case, as opposed to the previous case, where the images were learnt directly layer by layer, the states learnt are that of the 2-Dimensional Fourier Transform (2DFT) of the noisy images and a constant matrix, with all terms set to the mean of the noise.To execute this, we use 2 parallel single mode decoder layers, with the final learnt states being fed into the oracle that we introduce as $\hat{J}$  followed by a 2-Dimensional Inverse Fourier Transform (2DIFT). The oracle $\hat{J}$ essentially applies the displacement (difference) operator between the learnt states and acts as a quantum memory buffer [22] to store all the learnt states until they can be measured and the 2DIFT can be applied.

\begin{figure}
\centering
\includegraphics[width=180mm]{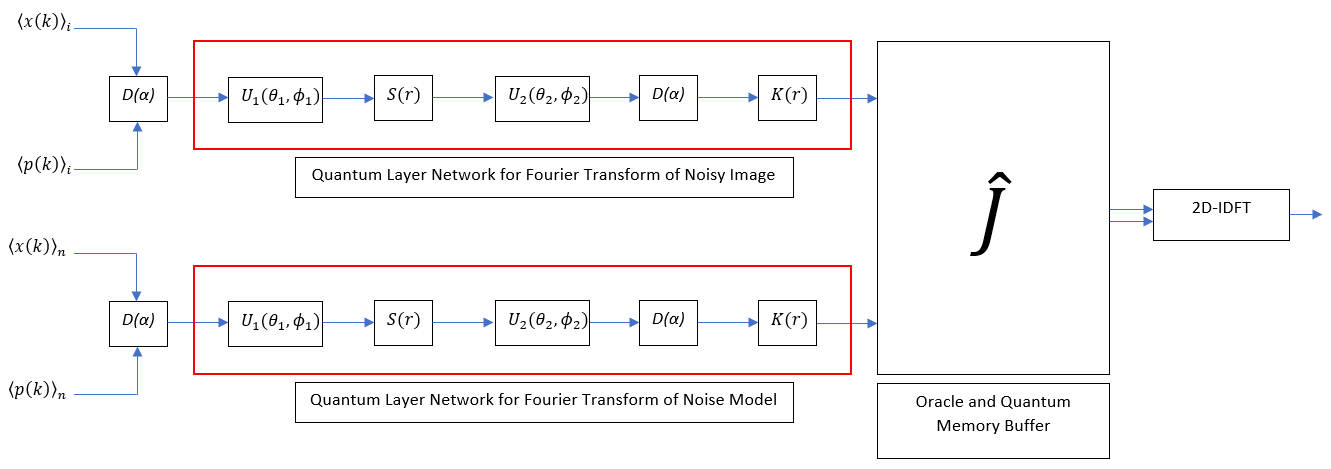}
\caption{\label{fig:epsart} Two parallel single mode quantum decoder layers, fed into the Oracle for storage and measurement - each learning the Fourier transforms of the noisy image and the noise
}
\end{figure}

Let $C_{i}$ represent the cost function for the noisy image and $C_{n}$ represent the cost function for the noise, with $\phi^{i}(k)$ and $\phi^{n} (k)$ representing the 2DFT of the noisy image and the noise respectively.
\begin{eqnarray*}
C_{i} = \Sigma _{j = 0}^{27} (|\braket{j}{\phi^{i}_{j}(k)}|^{2} - 1)^{2}\\
C_{n} = \Sigma _{j = 0}^{27} (|\braket{j}{\phi^{n}_{j}(k)}|^{2} - 1)^{2}
\end{eqnarray*}

Each layer is fed a single mode which has been encoded separately with either the noisy image or the noise, with the circuit diagram as shown above. Figure [5] illustrates the inputs and the outputs of the modified network.

\begin{figure}[h!]
\centering
\includegraphics{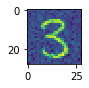}
\includegraphics{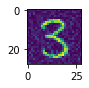}\\
\includegraphics{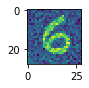}
\includegraphics{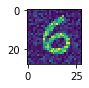}
\caption{\label{fig:epsart} Input and Output of the Denoising Quantum Decoder - the first row illustrates denoising the number "3" from the MNIST dataset with AWGN noise (mean = 0.5, standard deviation = 0.1) filtered to give a mean squared error of approximately 1\% while the second row illustrates the denoising of the number "6" from the MNIST dataset with AWGN noise (mean = 0.5, standard deviation = 0.2) filtered to give a mean squared error of approximately 4\%}
\end{figure}

We further ingeminate the flexibility of the network to learn any form of structured data with the example of coloured images - which are 2-dimensional arrays of pixel values, with each pixel containing the data of the red, green and blue colour intensities. Colored images can be trifurcated into their respective RGB values, each of which will be learnt separately via a 6-layer parallel single mode system. For the added purpose of demonstration, we have modified the architecture so as to present the denoised results of an image with the VIBGYOR colour spectrum.

\begin{figure}[hbt!]
\centering
\includegraphics{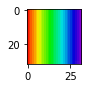}
\includegraphics{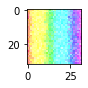}
\includegraphics{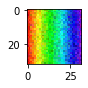}
\caption{\label{fig:epsart} Denoising a colorful image - the original image, the noisy image and the denoised image}
\end{figure}

An important caveat to note is that the AWGN noise is frequency independent, i.e. is a model that is mathematically tenable. While the network can be trained to identify frequency specific noise, filtering operations are yet to be explored.  

\section{Discussion} 

The proposed architecture has demonstrated exemplar performance for applications in image processing via deep neural networks. A key takeaway from this paper is the fact that no restrictions have been placed on the type of data being fed into the network or the type of encoding being used, implying that any form of structured data can be learnt by the network - either via multiple modes, or a single mode in the expanded basis. Furthermore, the this model is simple enough to be realized experimentally, with modifications to the architecture done by stacking a single layer, one after the other, depending upon the kind of data being processed. Moreover, the proposed architecture has demonstrated exemplar reconstruction and denoising capabilities using the quantum properties of light. Here, we see that a system as simple as a 6-layer single mode circuit could be used to denoise a coloured image with each layer learning it's RGB components and noise respectively. 

To better understand the performance of our network on the denoising, we plot the average mean squared error against the standard deviation of the noise added to the original figure. We observe an increasing correlation between the noisy and denoised image due to the fact that the AWGN model is immune to frequency selectivity and is used to provide a simple tractable model to visualize the working of the network before increasing the complexity of the noise added \cite{25}.

\begin{figure}
\centering
\includegraphics[width=90mm]{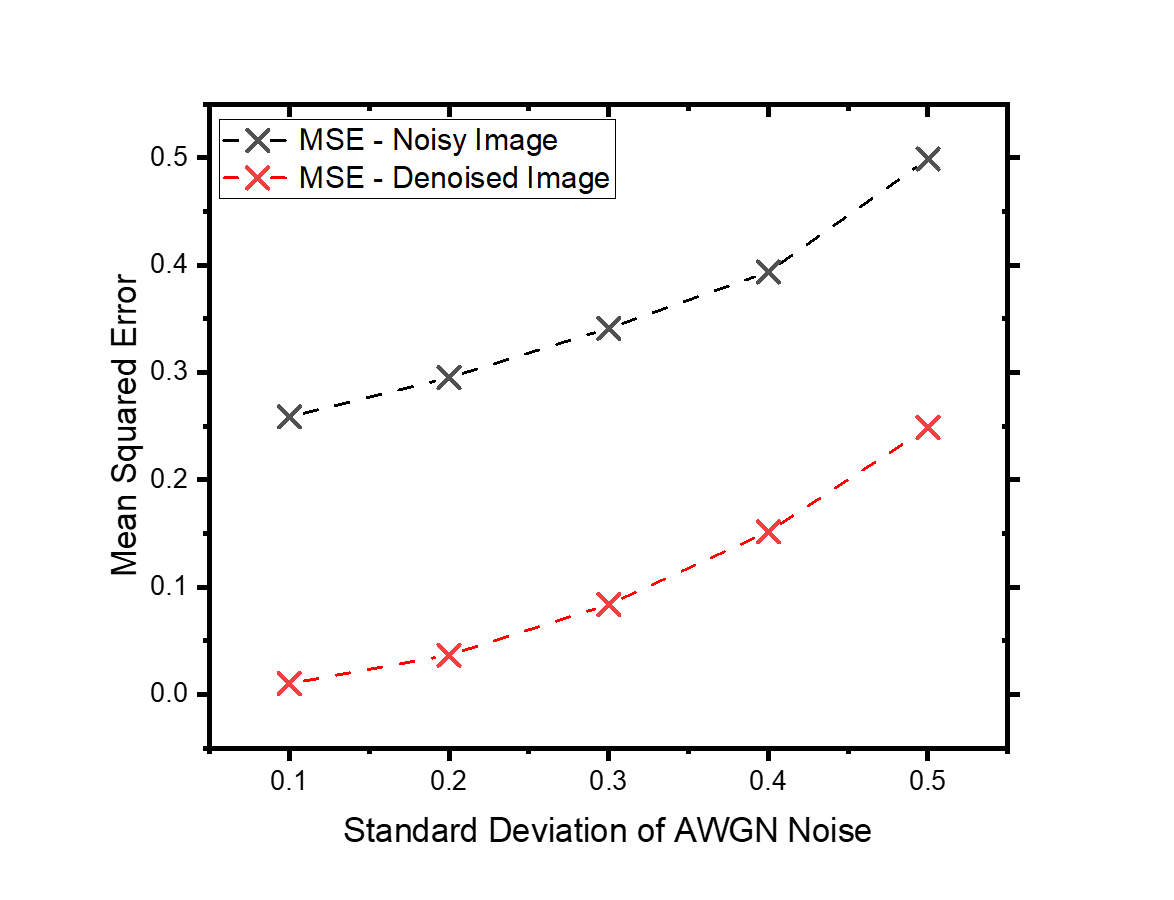}
\caption{\label{fig:epsart} Plot of Mean Squared Error against the standard deviation for the AWGN Noise for the noisy images and denoised images}
\end{figure}

The plot shows us that this architecture and network function well enough to remove noise approximately 30\% in mean squared error units. For higher values of standard deviation of the AWGN, the clarity of the image drops rapidly, rendering the reconstructed image useless. Experimental realizations of such a network would expose the photons to a multitude of sources of noise - while the Central Limit Theorem ascertains that the sum of independent noise variables would result in a Normal distribution, there would be an abundance of frequency dependant noise as well. We envision that such filtering components can be incorporated without altering the simplicity of the cost function. 

The network proposed uses a hybrid classical-quantum system - namely a classical network to feed the quantum network the appropriate structure of data. Such classical-to-quantum transfer of learnt parameters is perhaps the most appealing in today's era of NISQ devices \cite{32}. However, another viable model for transferring learnt parameters is the quantum-to-quantum transfer \cite{31} model, where instead of sourcing the computationally difficult calculations to the quantum section, the entire model (which could be computationally expensive due to lack of resources if done classically) is designed on purely quantum learning algorithms. Harnessing other properties of quantum mechanics - specifically, entanglement, would enable us to expand the basis of calculation and allow engineering a variety of other unexplored phenomena. Optical quantum information processing \cite{24} also presents with the added advantage that extremely large amounts of data can be handled at incredible speeds.

\section{Broader Impact and Conclusions}

Limitations in our nanofabrication technology, thereby our scaling ability and hence ending Moore's law has necessitated the new computational paradigm of quantum information processing which is becoming increasingly popular.  This has grown as field of interest with a plethora of engineering applications, giving us exponentially larger computational resources that scale with the number of qumodes being used. In this paper, we have demonstrated the use of continuous-variable model, as posited by the phase-space formalism of quantum mechanics to learn and process classical information, by encoding the same in the bosonic modes of an electomagnetic field. The flexibility of such a network is a vastly enabling tool with applications ranging from quantum cryptography to scalable quantum communication networks. We have shown via numerical experiments that such a network is extremely versatile in terms of the architecture - which can be modified by stacking the proposed experimental realization of the gates - as well as in terms of the format of the data being learnt. Our simulations have performed the classification, reconstruction and enhancement operations on the MNIST dataset of handwritten images. An important takeaway from the model we propound is the fact that there are no inherent restrictions on the type or structure of the data being learnt. Such as system would be of vast practical importance to all of our intelligent systems and communication networks.

Photons play an essential role in all quantum networking systems - either as information carriers \cite{26} or as mediators between quantum memories \cite{27}. Integration with spin based repeater nodes \cite{28} for the prospects of a quantum internet \cite{29} is a promising direction for future exploration. Another fruitful direction of research would be to amalgamate the other fundamental properties of quantum mechanics - specifically the uncertainty principle and entanglement to evaluate their role in determining the information capacity of a quantum neural network \cite{30}. We envision that such networks will be modified for a variety of other applications in signal processing, such as filtering or localized feature extraction. Furthermore, the development of an equivalent network to enable convolutional neural networks and hence allowing us to tackle a broader range of problems is an important  aspect of this model that remains to be explored.

While single mode networks and gates are yet to be realized to implement quantum machine learning algorithms and long distance communication efficiently, we hope that with the advent of emerging photonic technologies such as lossless fiber optics and quantum technologies like semiconductor based spin interaction dependant nodal systems, such systems will be integrated into our lives seamlessly.


\begin{thebibliography}{}

\bibitem{1}
N.R. Pal, S.K. Pal, A review on image segmentation techniques, Pattern Recognition 26 (1993) 1277–1294
\bibitem{2}
Knill, E., Laflamme, R. and Milburn, G. A scheme for efficient quantum computation with linear optics. Nature 409, 46–52 (2001). https://doi.org/10.1038/35051009
\bibitem{3}
Alejandro Perdomo-Ortiz, Marcello Benedetti, John Realpe-Gomez, and Rupak Biswas. Opportunities and challenges for quantum-assisted machine learning in near-term quantum computers. arXiv:1708.09757, 2017.
\bibitem{4}
Alessandro Ferraro, Stefano Olivares, and Matteo GA Paris. Gaussian states in continuous variable quantum information. arXiv preprint quant-ph/0503237, 2005.
\bibitem{5}
N. Killoran, T. R. Bromley, J. M. Arrazola, M. Schuld, N. Quesada, and S. Lloyd, “Continuous-variable quantum neural networks,” arXiv:1806.06871 (2018).
\bibitem{6}
R. Annavarapu, Singular Value Decomposition and the Centrality of Löwdin Orthogonalizations . American Journal of Computational and Applied Mathematics 3, 33 (2013).
\bibitem{7}
Deep Learning (Ian J. Goodfellow, Yoshua Bengio and Aaron Courville), MIT Press, 2016.
\bibitem{8}
Jun-ichi Yoshikawa, Shota Yokoyama, Toshiyuki Kaji, Chanond Sornphiphatphong, Yu Shiozawa, Kenzo Makino, and Akira Furusawa. Invited article: Generation of one-million-mode continuous-variable cluster state by unlimited time-domain multiplexing. APL Photonics, 1(6):060801, 2016.
\bibitem{9}
Ulrik L Andersen, Jonas S Neergaard-Nielsen, Peter Van Loock, and Akira Furusawa. Hybrid discrete and continuous-variable quantum information. Nature Physics, 11(9):713, 2015.
\bibitem{10}
DM Meekhof, C Monroe, BE King, Wayne M Itano, and David J Wineland. Generation of nonclassical motional states of a trapped atom. Physical Review Letters, 76(11):1796, 1996
\bibitem{11}
Chao Shen, Zhen Zhang, and L-M Duan. Scalable implementation of boson sampling with trapped ions. Physical review letters, 112(5):050504, 2014.
\bibitem{12}
K Moon and SM Girvin. Theory of microwave parametric down-conversion and squeezing using circuit QED. Physical review letters, 95(14):140504, 2005.
\bibitem{13}
Borja Peropadre, Gian Giacomo Guerreschi, Joonsuk Huh, and Alan Aspuru-Guzik. Proposal for microwave boson sampling. Physical review letters, 117(14):140505, 2016.
\bibitem{14}
M.G.A. Paris, Phys. Lett. A 217, Displacement Operator by Beam Splitter ; M. Ban, J. Mod. Opt. 44, 1175 (1997)
\bibitem{15}
Vahlbruch, H, M. Mehmet, K. Danzmann, and R. Schnabel (2016), “Detection of 15 db squeezed states of light and their application for the absolute calibration of photoelectric quantum efficiency,” Phys. Rev. Lett. 117, 110801
\bibitem{16}
Jun-ichi Yoshikawa, Toshiki Hayashi, Takayuki Akiyama, Nobuyuki Takei, Alexander Huck, Ulrik L. Andersen, and Akira Furusawa, “Demonstration of deterministic and high fidelity squeezing of quantum information,” Phys. Rev. A 76, 060301(R) (2007).
\bibitem{17}
LeCun, Y. and Cortes, C. , 'MNIST handwritten digit database' (2010)
\bibitem{18}
Nathan Killoran, Josh Izaac, Nicolás Quesada, Ville Bergholm, Matthew Amy, and Christian Weedbrook. “Strawberry Fields: A Software Platform for Photonic Quantum Computing”, Quantum, 3, 129 (2019).
\bibitem{19}
Thomas R. Bromley, Juan Miguel Arrazola, Soran Jahangiri, Josh Izaac, Nicolás Quesada, Alain Delgado Gran, Maria Schuld, Jeremy Swinarton, Zeid Zabaneh, and Nathan Killoran. “Applications of Near-Term Photonic Quantum Computers: Software and Algorithms”, arxiv:1912.07634 (2019).
\bibitem{20}
Martín Abadi, Ashish Agarwal, Paul Barham, Eugene Brevdo, Zhifeng Chen, Craig Citro, Greg S. Corrado, Andy Davis, Jeffrey Dean, Matthieu Devin, Sanjay Ghemawat, Ian Goodfellow, Andrew Harp, Geoffrey Irving, Michael Isard, Rafal Jozefowicz, Yangqing Jia, Lukasz Kaiser, Manjunath Kudlur, Josh Levenberg, Dan Mané, Mike Schuster, Rajat Monga, Sherry Moore, Derek Murray, Chris Olah, Jonathon Shlens, Benoit Steiner, Ilya Sutskever, Kunal Talwar, Paul Tucker, Vincent Vanhoucke, Vijay Vasudevan, Fernanda Viégas, Oriol Vinyals, Pete Warden, Martin Wattenberg, Martin Wicke, Yuan Yu, and Xiaoqiang Zheng. TensorFlow: Large-scale machine learning on heterogeneous systems, 2015. Software available from tensorflow.org.
\bibitem{21}
Diederik Kingma and Jimmy Ba. Adam: A method for stochastic optimization. In ICLR, 2015
\bibitem{22}
Hinton N. Srivastava and K. Swersky. Neural networks for machine learning-lecture 6a-overview of mini-batch gradient descent, 2012
\bibitem{23}
Bárány, Imre; Vu, Van (2007). "Central limit theorems for Gaussian polytopes". Annals of Probability. Institute of Mathematical Statistics. 35 (4): 1593–1621. arXiv:math/0610192. doi:10.1214/009117906000000791.
\bibitem{24}
Knill, E., R. Laflamme, and G. J. Milburn, 2001, Nature 409, 46. Kok, P., and B. Lovett, 2010, Introduction to optical quantum information processing (Cambridge University Press, Cambridge).
\bibitem{25}
Zhou, Y., Jiao, J., Huang, H., Wang, Y., Wang, J., Shi, H., Huang, T.: When AWGN-based denoiser meets real noises. arXiv preprint arXiv:1904.03485 (2019)
\bibitem{26}
Steinbrecher, G. R., J. P. Olson, D. Englund, and J. Carolan, “Quantum optical neural networks,” arxiv:1808.10047, 2018.
\bibitem{27}
A. I. Lvovsky, B. C. Sanders, and W. Tittel. Optical quantum memory. Nature Phot., 3:706, 2009.
\bibitem{28}
Atatüre, M., Englund, D., Vamivakas, N., Lee, S.-Y. and Wrachtrup, J. Material platforms for spin-based photonic quantum technologies. Nat. Rev. Mater. 3, 38–51 (2018).
\bibitem{29}
Kimble, H. J., “The quantum internet,” Nature (London) 453, 1023, 2008.
\bibitem{30}
L. G. Wright and P. L. McMahon, “The Capacity of Quantum Neural Networks,” arXiv:1908.01364 [quantph], Aug. 2019.
\bibitem{31}
A. Mari, T. R. Bromley, J. Izaac, M. Schuld, and N. Killoran, “Transfer learning in hybrid classical-quantum neural networks,” arXiv:1912.08278 [quant-ph, stat], Dec. 2019
\bibitem{32}
John Preskill, “Quantum Computing in the NISQ era and beyond,” (2018), arXiv:1801.00862.
\end{thebibliography}
\end{document}